\documentclass[conference]{IEEEtran}
\IEEEoverridecommandlockouts
\usepackage[style=numeric, sorting=none]{biblatex}
\usepackage{amsmath,amssymb,amsfonts}
\usepackage{algorithmic}
\usepackage{graphicx}
\usepackage{textcomp}
\usepackage{xcolor}
\usepackage{booktabs}
\usepackage{xurl}
\begin{document}

\title{Disentangled-Transformer: An Explainable End-to-End Automatic Speech Recognition Model with Speech Content-Context Separation\\

\thanks{The research was supported by the Flemish Government under “Onderzoeksprogramma AI Vlaanderen” and FWO-SBO grant S004923N: NELF.}
}

\author{\IEEEauthorblockN{1\textsuperscript{st} Pu Wang}
\IEEEauthorblockA{\textit{Department of Electrical Engineering-ESAT} \\
\textit{KU Leuven}\\
Leuven, Belgium \\
pu.wang@esat.kuleuven.be}
\and
\IEEEauthorblockN{2\textsuperscript{nd} Hugo Van hamme}
\IEEEauthorblockA{\textit{Department of Electrical Engineering-ESAT} \\
\textit{KU Leuven}\\
Leuven, Belgium \\
hugo.vanhamme@esat.kuleuven.be}
}


\maketitle

\begin{abstract}
End-to-end transformer-based automatic speech recognition (ASR) systems often capture multiple speech traits in their learned representations that are highly entangled, leading to a lack of interpretability. In this study, we propose the explainable Disentangled-Transformer, which disentangles the internal representations into sub-embeddings with explicit content and speaker traits based on varying temporal resolutions. Experimental results show that the proposed Disentangled-Transformer produces a clear speaker identity, separated from the speech content, for speaker diarization while improving ASR performance.
\end{abstract}

\begin{IEEEkeywords}
Explainable AI, Speech representation disentanglement, Speech recognition, Speaker diarization.
\end{IEEEkeywords}

\section{Introduction}
\label{sec:intro}
Modern automatic speech recognition (ASR) systems leverage data-driven deep learning methods. The emergence of end-to-end (E2E) encoder-decoder architectures, such as self-attention-based transformer models~\cite{vaswani2017attention, gulati2020conformer}, has markedly enhanced recognition accuracy, paving the way for advanced speech recognition applications like Google Home Assist, Apple Siri, and more. However, deep learning methods are often perceived as ``black boxes''.

In particular, transformer-based speech models often capture multiple speech traits in the learned representations at each encoder layer. These traits are highly entangled—not only the intended speech content (the actual linguistic information) but also speaker identity, dialect, accent, emotion, background noise, and other contextual factors~\cite{pasad2021layer, feng2022silence, chen2022large, chen2022does, chen2022wavlm}. For example, researchers have observed speaker characteristics and gender attributes in some of the learned representations of the speech foundation model Wav2Vec 2.0~\cite{fan2021exploring, choi2022opening}, leading to a lack of clear interpretability.

This lack of interpretability complicates the detection of flaws or biases within ASR models when errors occur due to data mismatches. Specifically, in the real world, speech data encompass significant diversity, including variations in speakers, such as healthy speakers versus dysarthric speakers, and speaking environments, for instance, from a single-speaker lecture to dialogues involving multiple speakers with overlapping speech. Research shows that training ASR models solely on the majority ``standard'' speakers-typically adult, first-language speakers without speech disabilities-results in notable disparities in recognition accuracy across different speaker groups with diverse sociolinguistic backgrounds~\cite{zhang2023exploring, garnerin2021investigating, zhang2022mitigating, kathania2020study, bartelds2023making}. For example, Wang et al.~\cite{wang2023benefits} show that speech foundation models fine-tuned on one group of dysarthric speakers exhibit significant fluctuations in recognition performance when tested on other dysarthric groups. While the linguistic information or speech content is expected to remain consistent across different speakers, the model produces speech content-related representations that are entangled with other speech traits not present in the training data, such as speaker identity. Furthermore, studies show that in well-trained transformer-based speech models, representations extracted from certain attention heads in the first several encoder layers form clear speaker and gender clusters, while other layers or attention heads do not exhibit the same behavior~\cite{pasad2021layer, feng2022silence, chen2022large, chen2022does, chen2022wavlm}. This arbitrary and inconsistent entanglement can occur at various components or layers within the model, making it challenging to answer the question ``\textit{Which} aspects of the learned representations contribute to fluctuations in recognition performance, and \textit{where} in the network architecture do they originate?''.
Consequently, this entanglement complicates efforts to improve the ASR model’s generalization.

\begin{figure*}[htbp]
\centerline{\includegraphics[width=1.0\linewidth]{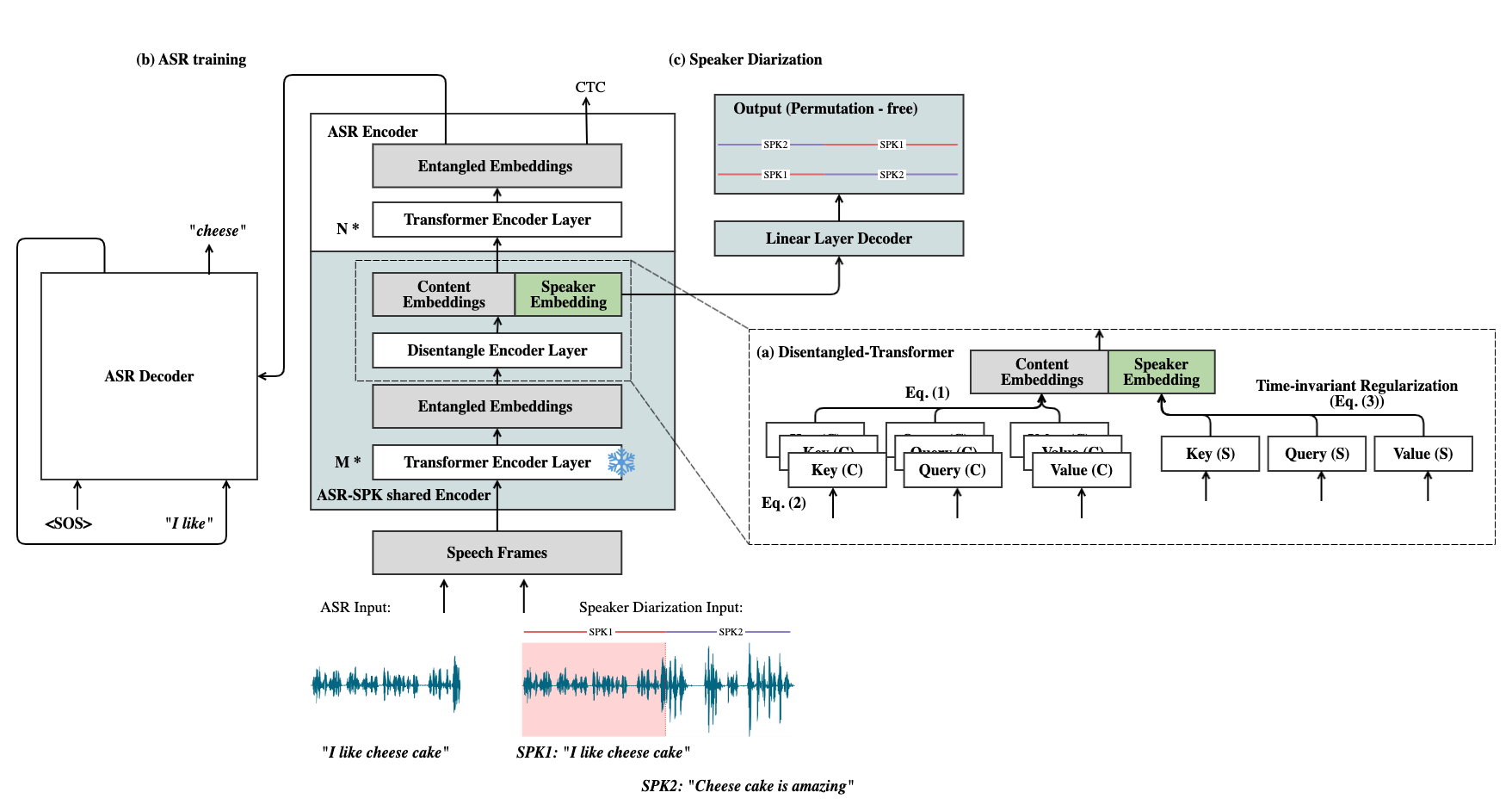}}
\caption{Structure of (a) Disentangled-Transformer; (b) E2E ASR model with Disentangled-Transformer (white rectangles); (c) E2E speaker diarization model with shared ASR Disentangled-Transformer encoder (blue rectangles).}
\label{fig:structure}
\end{figure*}

In this study, we focus on enhancing the explainability of learned representations in transformer-based E2E ASR models. We propose to disentangle the internal representation of the ASR encoder into sub-embeddings, with each sub-embedding explicitly correlated with a specific speech trait (such as speaker identity), without compromising ASR performance. 

In existing work, speech representation disentanglement has been primarily explored in speech editing or synthesis tasks. For instance, No{\'e} et al.~\cite{noe2021adversarial} proposed disentangling the sex attribute from the speaker representation to conceal personal information in a speech privacy preservation task. Yang et al.~\cite{yang2022speech} utilized mutual information learning to disentangle speaker information in a one-shot voice conversion (VC) task, enabling the conversion of speech content from one speaker to another's voice. Wang et al.~\cite{wang2023non} extended this approach to disentangle accent and speaker features in an accent transfer task, synthesizing speech with a specified accent. These methods are typically built using adversarial vector-quantized variational autoencoder (VQ-VAE) learning and reconstruction~\cite{noe2021adversarial, yang2022speech, wang2023non, tjandra2020transformer, lu2023speechtriplenet, polyak2021speech, williams2021learning, proszewska2022glowvc, liu2023disentangling}, which often come at the cost of reduced speech content resolution and unstable training. Considering that the main objective of this work is to enhance speech recognition, we avoid incorporating additional signal reconstructors or training efforts. Instead, we propose a simple yet effective approach to disentangle speech representations within an ASR task.

In vanilla transformers, the self-attention operation efficiently attends to all information across the full sequence simultaneously. However, useful information changes at different rates: the \textit{what} (linguistic content) requires a resolution of a few tens of milliseconds, while the \textit{who} (speaker identity, dialect, foreign accent) and \textit{how} (speaking rate, formality, stress) vary at a slower pace.  
Therefore, in this study, we propose to disentangle speech trait-specific sub-embeddings based on the different temporal behaviors of each trait. Specifically, we build a Disentangled-Transformer with explicit content-embedding attention heads that capture the rapidly varying embeddings, and a speaker-embedding attention head that captures the slowly varying embeddings. This is implemented by introducing time-invariant regularization to penalize rapid changes in the speaker-embedding attention head of the encoder layer during ASR training. This regularization can be applied to either a single layer or the full set of encoder layers. To assess its explainability with respect to disentangled speaker embedding, the Disentangled-Transformer encoder is further shared with an E2E speaker diarization task by adding a single linear decoder layer. It is evaluated under conditions including single speaker and noise separation, two-speaker non-overlapping diarization, and two-speaker fully-overlapping separation.

In Section~\ref{sec:method}, we first recapitulate the transformer architecture and introduce the Disentangled-Transformer. The implementations for ASR and speaker diarization are also detailed in this section. Section~\ref{sec:experiment} discusses the experimental setup and datasets used for evaluation, and the corresponding results are presented in Section~\ref{sec:results}. In Section~\ref{sec:conclusion} we will conclude our work.

\section{Method}
\label{sec:method}
\subsection{Transformer encoder}
\label{sec:transformer}
The transformer encoder is a layer-stacked model with each layer containing one multi-head self-attention (MHSA) block and one feed-forward block~\cite{vaswani2017attention}. The output of the MHSA for the $i$-th frame is a linear projection of the concatenated multi-scaled dot-product self-attention operations as shown in~(\ref{eq:MHSA})
\begin{equation}
  {MHSA}_i = W_{out}Concat(Attn^1_i,Attn^2_i,...Attn^N_i)
  \label{eq:MHSA}
\end{equation}

In each of the $N$ attention heads, the feature representation of the sequential data is first linearly transformed into a sequence of Keys, Values and Queries. The feature representation in the next layer is built as a non-linear mapping of a weighted average of the Values. The weight of each Value in the $n^{th}$ attention head is determined by the similarity between the Key and the Query, as measured by a dot-product~(\ref{eq:v_trans}). 
\begin{equation}
  Attn^n_i = \sum_j S^n_{ij} V^n x_j \text{ with } S^n_{ij} = \text{Softmax}\left(\frac{x_j^T{K^n}^T Q^nx_i}{\sqrt{d_k^n}} \right)
  \label{eq:v_trans}
\end{equation}
where $x$ is the $d$-dimensional feature embedding of the previous layer. $Q^n, K^n\in\mathbb{R}^{d_k^n \times d}$, $V^n\in\mathbb{R}^{d_v^n \times d}$ are the projection matrices for Query, Key and Value in the $n^{th}$ attention head respectively. $d_k^n$ is the width of the $n^{th}$ attention head.

\subsection{Disentangled-Transformer}
\label{sec:disen-transformer}
The idea of disentanglement is to partition the embedding from one encoder layer into sub-embeddings with different temporal behaviour. 

As shown in Figure~\ref{fig:structure} (a), the embeddings from the previous encoder layer are disentangled into content embeddings $c$ and speaker embeddings $s$ using separate Query, Key and Value projection matrices for content related weights and speaker related weights. As explained in Section~\ref{sec:intro}, content embeddings change rapidly over time, while speaker embeddings remain stable for the same speaker. To achieve low temporal variation in speaker embeddings $s$, time-invariant regularization terms for $s$ in each time frame $t$ are added to the total loss during training with the penalty scale $\lambda_{s}$ ($\lambda_{s} = 0.1$ is used by default):
\begin{equation}
  L_{s} = \lambda_{s} \frac{1}{L} \sum_l \frac{1}{\sqrt{d_{s}}} 
  \sum_t (\sqrt{||s_{t+1} - s_t||^2} + \sqrt{||s_{t+5} - s_t||^2})
  \label{eq:disen_trans}
\end{equation}
where, $l$ indicates the index of the constrained layer (or Disentangled-Transformer layer) within the transformer encoder, $d_s$ is the dimension of the speaker embedding, and $s_t$ is the speaker embedding at the $t^{th}$ frame. 

With the constraint $\sqrt{||s_{t+5} - s_t||^2}$, the speaker embeddings are encouraged to remain stable over 5 frames (typically 25 milliseconds per frame with a 10-millisecond frame shift), which is longer than the average time resolution for uttering a phoneme. An additional constraint with a time step 1 $\sqrt{||s_{t+1} - s_t||^2}$ is added to prevent periodic noise. Here, we apply the constraint to a single attention head of the Disentangled-Transformer layer, while the rest of the attention heads in the same layer serve as content embeddings. The Disentangled-Transformer layer can be applied to the single layer or the full set of layers in the transformer encoder.

\subsection{ASR with Disentangled-Transformer}
\label{sec:asr}
The white rectangles in Figure~\ref{fig:structure} (b) illustrate the E2E ASR model, which is built with stacked transformer layers and Disentangled-Transformer layers as the encoder, and stacked transformer layers as the decoder. It takes utterances from the single speaker as the inputs and generates transcripts as the outputs.

The ASR model adopts a hybrid CTC/attention architecture from ESPnet~\cite{watanabe2017hybrid, watanabe2018espnet}, training with a combination of CTC loss $L_{ctc}$, attention-based cross-entropy loss $L_{attn}$ and time-invariant regularization $L_{s}$:
\begin{equation}
  L_{asr} = \alpha L_{ctc} + (1-\alpha) L_{attn} + L_{s}
  \label{eq:asr}
\end{equation}
We use $\alpha=0.3$ in this study. During decoding, CTC scores are combined in one-pass beam search to further eliminate irregular alignments.

\subsection{Speaker diarization with Disentangled-Transformer}
\label{sec:sd}
The blue rectangles in Figure~\ref{fig:structure} (c) illustrate the E2E speaker diarization model. It takes utterances from at least one speaker as input and indicates \textit{who} speaks \textit{when}.

The main target of speaker diarization task is to assess the disentangled sub-embeddings. Therefore, it shares the Dientangle-Transformer encoder with the ASR model and uses speaker embeddings $s$ to estimate speaker(s)' activities with a linear decoder layer.  The speaker diarization model predicts speaker activity as binary multi-class labels $\in{\{0,1\}}^{num_{spk} \times T}$, where $num_{spk}$ is the number of speakers. Therefore, it can model overlapping speech by making the labels of overlapping speech frames all active. 

Both training and decoding are conducted in a permutation-free manner~\cite{fujita2019end, maiti2023eend}. As shown in the ``Output'' rectangle in Figure~\ref{fig:structure} (c). The speaker label itself is not the focus, rather, detecting speech activity or speaker switches is more important.

\section{Experimental Setup}
\label{sec:experiment}
\subsection{Dataset}
\label{sec:Dataset}
\textit{LibriSpeech100h} is used for training the Disentangled-Transformer ASR model. It contains 100 hours of read English speech derived from audiobooks~\cite{panayotov2015librispeech}. Each recording of LibriSpeech100h is utterance from the single speaker. The ASR model is validated on dev and test splits: \textit{dev\underline{~}clean}, \textit{dev\underline{~}other}, \textit{test\underline{~}clean} and \textit{test\underline{~}other}, they are named based on the difficulty ASR systems may face in processing them.  Each of the dev and test sets contains about 5 hours of audio.

\textit{LibriMix} is used for training and validating the Disentangled-Transformer speaker diarization model. Each audio in LibriMix data consists of one- or two-speaker mixture(s), with or without noise. The speech utterances in LibriMix are taken from LibriSpeech100h, dev, and test splits, and the noise samples are from WHAM!~\cite{cosentino2020librimix}. During the evaluation phase, to avoid background noise in the reference signals,
we only use the speech from \textit{dev\underline{~}clean}, and 
\textit{test\underline{~}clean}.

\begin{figure}[htbp]
\centerline{\includegraphics[width=1.0\linewidth]{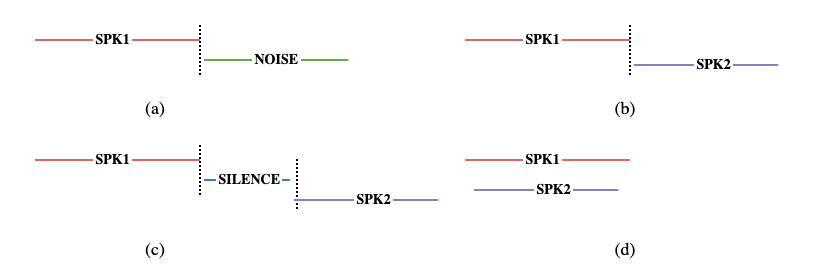}}
\caption{LibriMix example of: (a) LibriMix 1.0: single speaker’s utterance, non-overlapping, mixed with WHAM! noise; 
(b) LibriMix 2.0: two speakers’ utterances, non-overlapping, no-silence mixture; 
(c) LibriMix 3.0: two speakers’ utterances, non-overlapping, with silence in the mixture; and (d) LibriMix 4.0: two speakers’ utterances, fully overlapping. }
\label{fig:librimix}
\end{figure}

To stimulate realistic, conversation-like scenarios, we establish four versions of datasets: 
(a) LibriMix 1.0: single speaker’s utterances, non-overlapping, mixed with WHAM! noise; 
(b) LibriMix 2.0: two speakers’ utterances, non-overlapping, no-silence mixture; 
(c) LibriMix 3.0: two speakers’ utterances, non-overlapping, with silence in the mixture; and (d) LibriMix 4.0: two speakers’ utterances, fully overlapping. 
Examples from these four datasets are shown in Figure~\ref{fig:librimix}. Mixtures of LibriMix 2.0, 3.0, and 4.0 are generated by randomly selecting utterances for different speakers. The loudness of each utterance is uniformly sampled between -25 and -33 LUFS.

\subsection{Implementation details}
\label{sec:implementation}
Both the ASR model and speaker diarization model are implemented with ESPnet. 

The ASR model is build with an 18-layer (Disentangled-)Transformer encoder and a 6-layer transformer decoder. Each layer consists of a 4-head 64-dimensional 
MHSA and a feed-forward layer with the inner dimensions of 1024. The Disentangled-Transformer layer(s) replaces one or multiple transformer encoder layer(s) within the model. In the Disentangled-Transformer layer, time-invariant regularization is applied to the $4^{th}$ attention head as speaker embeddings $s$, while the remaining three attention heads are used for content embeddings $c$. The model is trained with default configurations specified in the \textit{librispeech\underline{~}100h} recipe\footnote{\url{https://github.com/espnet/espnet/tree/master/egs2/librispeech_100/asr1}.}. 

The speaker diarization model is built with the shared ASR disentangled encoder and a linear decoder layer. During training, all encoder parameters are frozen except for the Dientangled-Transformer layer. The default training configuration can be found in the \textit{librimix} recipe\footnote{\url{https://github.com/espnet/espnet/tree/master/egs2/librimix/diar1}.}. 

\subsection{Evaluation metrics}
\label{sec:evaluation}
We report ASR performance using word error rate (WER(\%)) and diarization
performance using diarization error rate (DER(\%))~\cite{fujita2019end, maiti2023eend} and speaker error time (the total duration of incorrectly attributed speech in seconds). When calculating the DER, collar tolerance of 0.0 sec and median filtering
of 11 frames are applied. 

\section{Experimental Results}
\label{sec:results}
\subsection{ASR performance}
As explained in Section~\ref{sec:intro} and Section~\ref{sec:implementation}, the goal of this study is to build an ASR model with explicitly disentangled speaker embeddings without hurting its original ASR performance. Therefore, we use the standard benchmark transformer ASR model which is built with an 18-layer transformer encoder and a 6-layer transformer decoder, as the baseline. 

\begin{figure}[htbp]
\centerline{\includegraphics[width=1.0\linewidth]{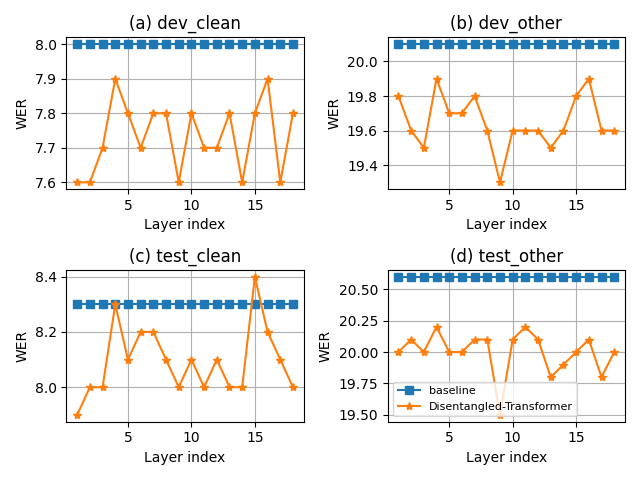}}
\caption{WER (\%) as a function of indexed layers for the baseline transformer (blue) and the proposed Disentangled-Transformer (orange) on (a) \textit{dev\underline{~}clean}; (b) \textit{dev\underline{~}other}; (c) \textit{test\underline{~}clean}; (d) \textit{test\underline{~}other}, trained on the Librispeech100h}
\label{fig:asr}
\end{figure}

We first substitute each transformer encoder layer of the benchmark model with the Disentangled-Transformer layer, starting from the bottom to the top. The WER results of each model are shown in Figure~\ref{fig:asr}, where the x-axis represents the layer index that is replaced with the Disentangled-Transformer and the y-axis represents the WER.   

In Figure~\ref{fig:asr}, the blue dots record the benchmark results, and the orange dots record the Disentangled-Transformer results. 
All Disentangled-Transformer models significantly outperform the benchmark model, except when replacing the $15^{th}$ encoder layer with Disentangled-Transformer on \textit{test\underline{~}clean}, which led to 0.1\% performance degradation. A potential reason for this, as suggested in~\cite{champion2022disentangled, adi2019reverse}, is that representations extracted from deeper layers in networks are less likely to encode speaker information. This implies that the upper layers, such as the $15^{th}$ encoder layer in this study, are designed to focus more on spoken content and potentially discard speaker information. While with a large penalty scale $\lambda_{s}$, more speaker information is retained, which reduces the resolution of spoken content. Therefore, a smaller penalty scale $\lambda_{s}$ is expected to be more suitable for the upper layers when substituting a single Disentangled-Transformer layer.


An alternative approach is to substitute multiple layers. By using the average regularization loss $L_s$ across all replacement layers in Eq.~(\ref{eq:disen_trans}), the model can dynamically adjust the penalty distribution to more suitable layers. Therefore, we further replace all the encoder layers with Disentangled-Transformer. The WER results are summarized in Table~\ref{tab:asr}. The results show that the proposed disentanglement operation does not degrade the performance compared to the baseline. On the contrary, the model benefits from the regularization and shows improvement.

\begin{figure*}[htb]
\centerline{\includegraphics[width=0.7\linewidth]{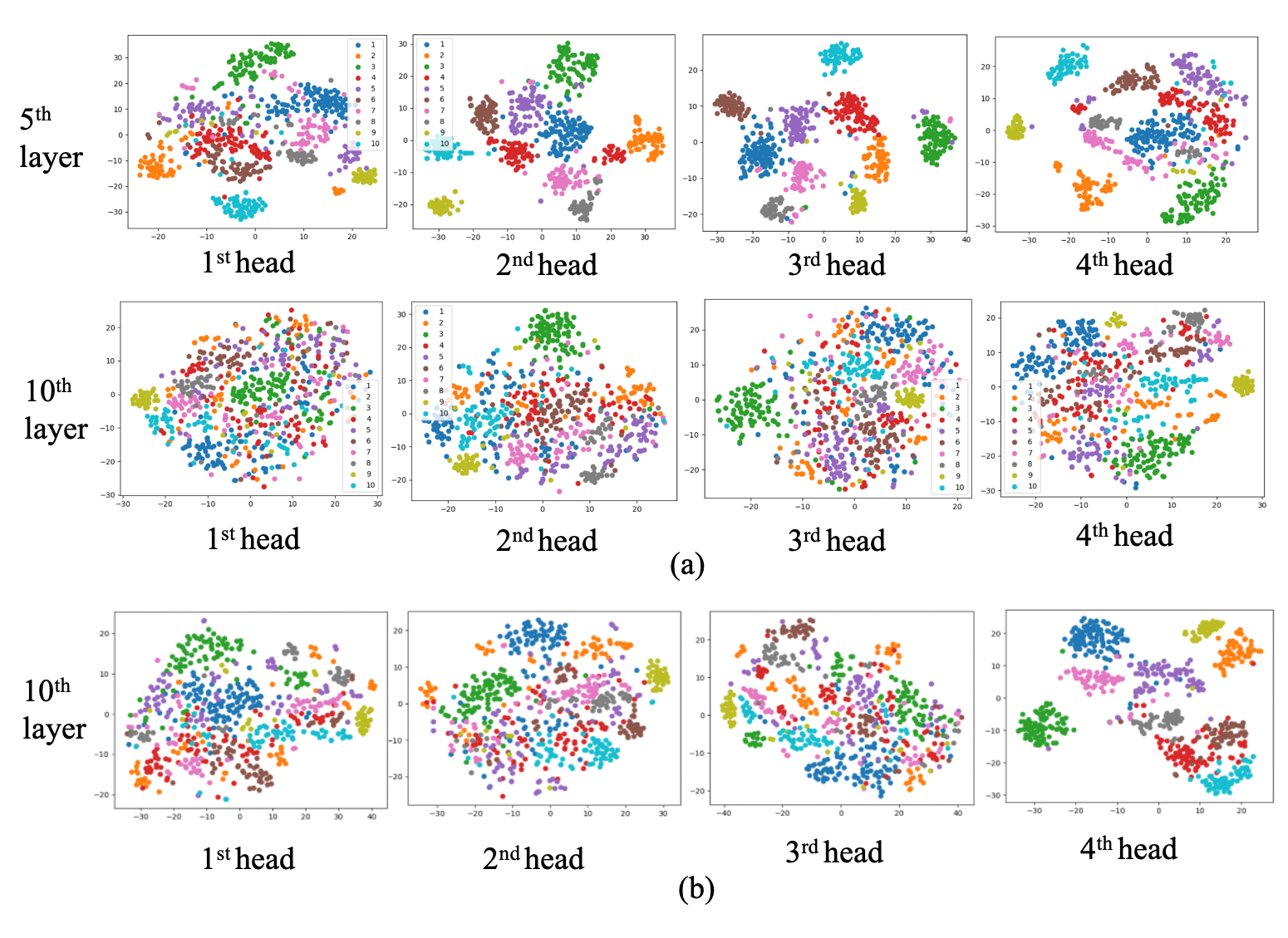}}
\caption{T-SNE plots of embeddings extracted from each attention head of different encoder layers of (a)baseline transformer and (b)Disentangled-Transformer.}
\label{fig:t-sne-perhead}
\end{figure*}

\begin{table}[htb]
\centering
\label{tab:asr}
\caption{WER (\%) for the baseline transformer and Disentangled-Transformer, trained on Librispeech100h}
\scalebox{0.9}{
\begin{tabular}{ccccc}
\toprule
 Method & dev\underline{~}clean & dev\underline{~}other & test\underline{~}clean & test\underline{~}other \\
\hline
Baseline transformer& $8.0$ & $20.1$ & $8.3$ & $20.6$ \\
Disentangled-Transformer & $7.8$ & $19.6$ & $8.1$ & $20.0$ \\
\bottomrule
\end{tabular}
}
\end{table}

\begin{figure*}[htb]
\centerline{\includegraphics[width=0.7\linewidth]{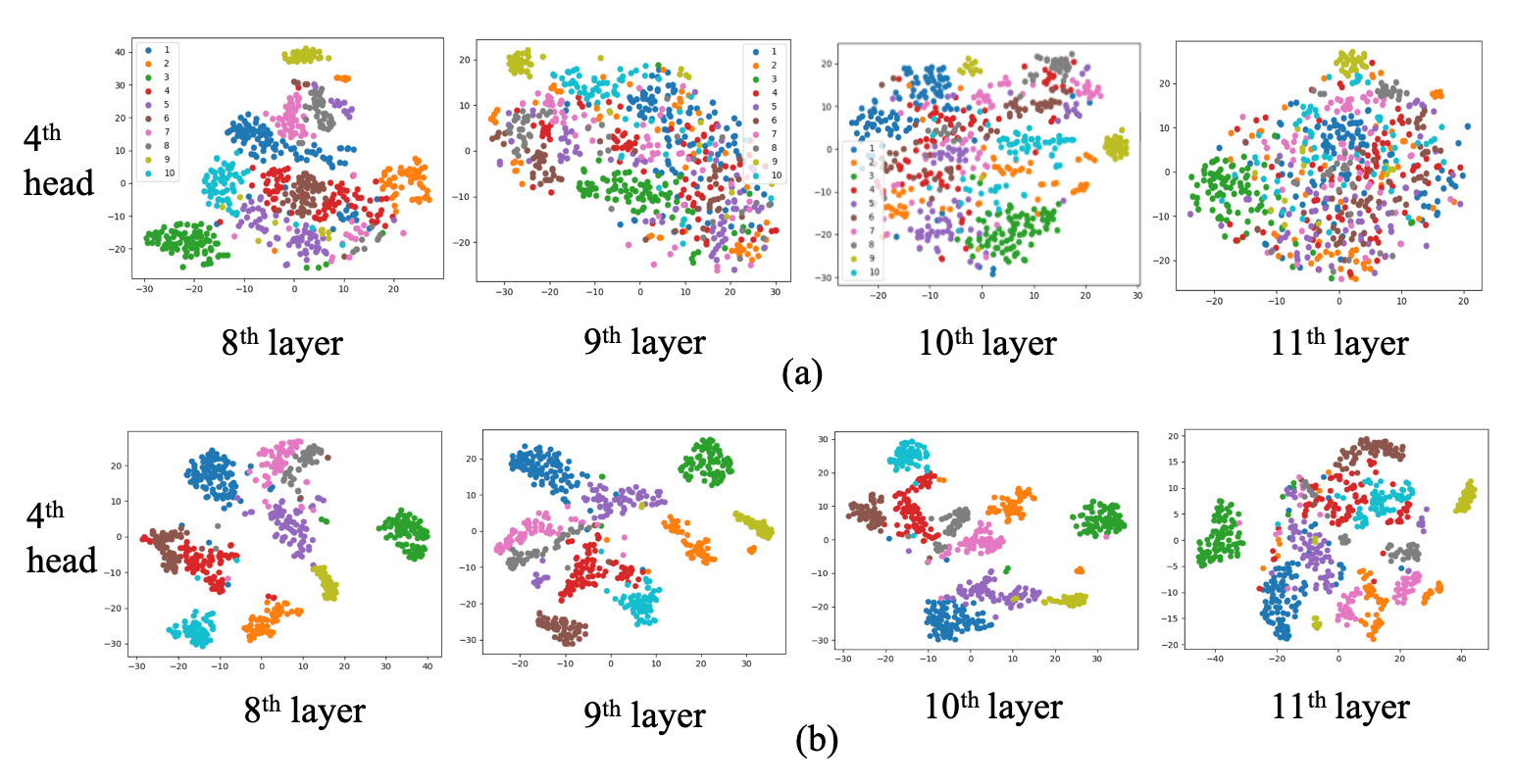}}
\caption{T-SNE plots of embeddings extracted from the $4^{th}$ attention head of (a)baseline transformer and (b)Disentangled-Transformer.}
\label{fig:t-sne-4head}
\end{figure*}

To visualize the baseline entangled embeddings and disentangled embeddings, we randomly select 8 minutes of speech from 10 speakers in the \textit{test\underline{~}clean} dataset and obtain the learned representations from each attention head across different layers from both the baseline transformer model and the Disentangled-Transformer model. The T-SNE polts are shown in Figure~\ref{fig:t-sne-perhead}, where each color represents one speaker. A clustering pattern in the same color indicates clear speaker clustering. 

From Figure~\ref{fig:t-sne-perhead} (a), we observe that in the baseline transformer, not every attention head contains clear speaker information. For example, the $3^{nd}$ attention head in the $5^{th}$ layer shows a clear speaker pattern while different colors in the $10^{th}$ layer blend together. This aligns with Section~\ref{sec:intro} that when we extract learned representations from arbitrary layers and attention heads, it is difficult to determine which specific speech traits are present in the extracted representations, leading to a lack of explainability. In contrast, after applying the time-invariant constraint to the $4^{th}$ attention head of the Disentangled-Transformer layer, the extracted embeddings, as shown in Figure~\ref{fig:t-sne-perhead} (b) ``$10^{th}$ layer $4^{th}$ head'', consistently show significant speaker patterns.

This is further verified by comparing the embeddings extracted from the $4^{th}$ attention head in the $8^{th}$ to $11^{th}$ layers of both the baseline transformer and the Disentangled-Transformer model. The T-SNE results of 10 speakers are shown in Figure~\ref{fig:t-sne-4head}, where the Disentangled-Transformer always shows clear speaker patterns, even in the higher layers, which can serve as speaker embeddings. In contrast, the baseline transformer model does not exhibit clear speaker information. 

\subsection{Speaker diarization performance}
To further quantitatively assess the explainability of the disentangled embeddings, the disentangled encoder trained on ASR targets is shared to build a speaker diarization model for the Librimix dataset. The benchmark model built in ESPnet is configured with 4 transformer encoder layers~\cite{fujita2019end}, and we adopt a similar configuration, using 3 transformer encoder layers and 1 Disentangled-Transformer layer at the top. Both models are followed by a linear decoder layer trained for 5, and 10 epochs. The first 3 transformer encoder layers in the Disentangled-Transformer are frozen during training, since the Disentangled-Transformer can be seen as a pre-trained model (as all speakers in LibriMix are also present in LibriSpeech100h). For a fair comparison, we train a transformer ASR model on Librispeech100h without the time-invariant constraint and use the same embeddings as the Disentangled-Transformer for the LibriMix speaker diarization, we refer it as ``ASR-transformer'' model. The performances on LibriMix 1.0-4.0 are shown in Table~\ref{tab:spk_v1},\ref{tab:spk_v2},\ref{tab:spk_v3},\ref{tab:spk_v4}.
\begin{table}[htb]
\centering 
\caption{DER (\%) for the Benchmark transformer, Disentangled-Transformer, and ASR-transformer on LibriMix 1.0 over 5 and 10 epochs}
\label{tab:spk_v1}
\scalebox{0.9}{
\begin{tabular}{lcccc} 
\toprule
 Method & \multicolumn{2}{c}{5 epochs} & \multicolumn{2}{c}{10 epochs} \\
  & dev\_clean & test\_clean & dev\_clean & test\_clean \\
\midrule
Benchmark Transformer & $19.5$ & $18.2$ & $7.6$ & $8.2$ \\
Disentangled-Transformer & $14.5$ & $12.7$ & $5.9$ & $5.7$ \\
ASR-transformer & $19.8$ & $17.7$ & $7.5$ & $7.9$ \\
\bottomrule
\end{tabular}
}
\end{table}
\begin{table}[htbp]
\centering 
\caption{Speaker error time (sec) in total 28618.22 seconds speech time for the Benchmark transformer, Disentangled-Transformer, and ASR-transformer trained on LibriMix 2.0 over 5 and 10 epochs}
\label{tab:spk_v2}
\scalebox{0.9}{
\begin{tabular}{lcccc} 
\toprule
 Method & \multicolumn{2}{c}{5 epochs} & \multicolumn{2}{c}{10 epochs} \\
  & dev\_clean & test\_clean & dev\_clean & test\_clean \\
\midrule
Benchmark Transformer & $2575.6$ & $2489.8$ & $77.9$ & $73.8$ \\
Disentangled-Transformer & \textbf{$1631.4$} & \textbf{$1602.6$} & \textbf{$6.5$} & \textbf{$6.9$} \\
ASR-transformer & $2692.4$ & $3033.5$ & $92.7$ & $86.2$ \\
\bottomrule
\end{tabular}
}
\end{table}
\begin{table}[htb]
\centering 
\caption{DER (\%) for the Benchmark transformer, Disentangled-Transformer, and ASR-transformer trained on LibriMix 3.0 over 5 and 10 epochs}
\label{tab:spk_v3}
\scalebox{0.9}{
\begin{tabular}{lcccc} 
\toprule
 Method & \multicolumn{2}{c}{5 epochs} & \multicolumn{2}{c}{10 epochs} \\
  & dev\_clean & test\_clean & dev\_clean & test\_clean \\
\midrule
Benchmark Transformer & $16.87$ & $18.56$ & $7.05$ & $7.31$ \\
Disentangled-Transformer & $7.9$ & $8.7$ & $2.6$ & $2.5$ \\
ASR-transformer & $20.4$ &$ 23.3$ & $8.8$ & $8.5$ \\
\bottomrule
\end{tabular}
}
\end{table}
\begin{table}[htb]
\centering 
\caption{DER (\%) for the Benchmark transformer, Disentangled-Transformer, and ASR-transformer trained on LibriMix 4.0 over 5 and 10 epochs}
\label{tab:spk_v4}
\scalebox{0.9}{
\begin{tabular}{lcccc} 
\toprule
 Method & \multicolumn{2}{c}{5 epochs} & \multicolumn{2}{c}{10 epochs} \\
  & dev\_clean & test\_clean & dev\_clean & test\_clean \\
\midrule
Benchmark Transformer & $19.2$ & $19.4$ & $9.8$ & $9.0$ \\
Disentangled-Transformer & $10.7$ & $10.6$ & $5.6$ & $5.6$ \\
ASR-transformer & $18.5$ & $21.6$ & $13.3$ & $12.9$ \\
\bottomrule
\end{tabular}
}
\end{table}

From these Tables, Disentangled-Transformer significantly outperforms both the benchmark transformer and the ASR-transformer, thanks to the explicit speaker embeddings. Pre-training on ASR target does not help the ASR-Transformer to achieve better performance compared to the benchmark transformer.
\section{Conclusion}
\label{sec:conclusion}
End-to-end encoder-decoder-based ASR models lack interpretability due to the entangled representations captured in each encoder layer. We therefore propose disentangling the internal representations into explicit content embeddings and speaker embeddings based on the different temporal behaviors of these two speech traits. Experimental results on the speaker diarization task show that the Disentangled-Transformer-based ASR model is more interpretable, with clear speaker identity captured in a specific disentangled attention head.

\section*{Acknowledgment}

The research was supported by the Flemish Government under “Onderzoeksprogramma AI Vlaanderen” and FWO-SBO grant S004923N: NELF.

\printbibliography

\end{document}